\documentclass[prl,twocolumn,superscriptaddress]{revtex4}
\usepackage{graphicx}
\usepackage{amsmath}

\begin{document}

\title{Noisy traveling waves: effect of selection on genealogies}

\author{E. Brunet}
\author{B. Derrida}
\affiliation{Laboratoire de Physique Statistique, \'Ecole Normale
Sup\'erieure,
24 rue Lhomond, 75231 Paris cedex 05, France}
\author{A.~H. Mueller}
\affiliation{Department of Physics, Columbia University,
New York, NY 10027, USA}
\author{S. Munier}
\affiliation{Centre de Physique Th{\'e}orique, Unit\'e mixte de
recherche du CNRS (UMR 7644),
{\'E}cole Polytechnique, 91128~Palaiseau, France}
\pacs{02.50.-r, 05.40.-a, 89.75.Hc}

\begin{abstract}
For a family of models of evolving population under selection, which can
be described by noisy traveling wave equations, the coalescence times
along the genealogical tree scale like $\ln^\alpha N$, where $N$ is the
size of the population, in contrast with neutral models for which they
scale like $N$. An argument relating
this time scale to the diffusion constant of the noisy traveling wave
leads to a prediction for $\alpha$ which agrees with our simulations.
An exactly soluble case gives
trees with statistics identical to those predicted for mean-field
spin glasses by Parisi's theory.
\end{abstract}

\maketitle
Traveling wave equations such as the F-KPP equation $\partial_t h=
\partial_{x}^2 h+ h
- h^2$  \cite{Fisher.37,KPP.37,vanSaarloos.03} describe how a stable medium $h=1$ invades
an unstable medium $h=0$. They were first introduced to study how an
advantageous gene propagates through a population, $h$ being the fraction
of the population with the advantageous gene. They also appear in other
contexts such as disordered
systems \cite{DerridaSpohn.88,CarpentierLedoussal.00,BrunetDerrida.04},
QCD \cite{IancuMuellerMunier.05},
reaction-diffusion \cite{Breuer.95,DoeringMuellerSmereka.03},
fragmentation \cite{KrapivskyMajumdar.00} or
chemistry \cite{Lemarchand.95}.

Traveling wave equations often represent a mean field picture where the
fluctuations at the microscopic scale are ignored. The effect of these
fluctuations can be
represented \cite{MuellerSowers.95,Mai.96,Lemarchand.95,DoeringMuellerSmereka.03,Panja.03}
by a noise term ($\partial_t
h=\partial_{x}^2 h + h - h^2 + \epsilon \eta(x,t) \sqrt{h-h^2}$).
Determining quantitatively the effect of a weak noise ($\epsilon \ll 1$)
on the front position is a subject of active research. There is increasing evidence that the
dynamics of the position of the front is dominated by the fluctuations
near its tip \cite{BrunetDerrida.01,BDMM.06,Kloster.05} and that there is
a shift in the velocity of the
front \cite{Breuer.95,Kessler.98,ColonDoering.05,Moro.04,PechenikLevine.99,Escudero.04}, logarithmic in the
amplitude $\epsilon$ of the noise, as predicted by a simple cut-off
theory \cite{BrunetDerrida.97}.

In the present letter, we consider models of an evolving population under
selection, which can be described by noisy traveling wave equations.
Instead of focusing on the time dependence of the position of the front,
we look at the problem from a different perspective: we determine how
coalescence times in the genealogy depend on the size of the
population.
Our simulations as well as a simple argument indicate that these
coalescence times scale as the inverse of the diffusion constant of the
front.

We consider a population of fixed size $N$ with asexual reproduction.
Each individual~$i$ is characterized by a real number $x_i$ 
measuring its adequacy to the environment, and the population is
fully specified by these $N$ positions $x_i$'s on the adequacy axis.
At each new generation, all
the individuals disappear after giving birth to some offsprings. We
consider the following variants:

In \textbf{Model~A}, each individual gives birth to $k$ offsprings, and
the $j$-th offspring of individual~$i$ is at position $x_i +
\epsilon_{i,j}$, where the $\epsilon_{i,j}$ are uncorrelated random
numbers chosen according to some distribution $\rho(\epsilon)$. Thus,
each individual inherits its parent's adequacy, and
$\epsilon_{i,j}$ accounts for the effects of mutation. Then comes the
selection step: out of the $kN$ new individuals, we only keep the $N$
best ones, the ones with the highest~$x_i$'s. Typically, we will take
$k=2$ and $\rho(\epsilon)$ uniform between~0 and~1. 
A similar model was proposed
recently \cite{PengGerlandHwaLevine.03,Snyder.03,KlosterTang.04,Kloster.05}
to study the evolution under competitive selection of a
population of DNA molecules in vitro. The population undergoes several
cycles where, in the first part of a cycle, each molecule is amplified
by a fixed number $k$ with possible mutations and in the second part
of the cycle selection acts by keeping only the best $1/k$ fraction of
these offsprings, defined as the molecules with the highest binding
energies to a given target. In this picture, $x_i$ represents this
binding energy. 

We also investigate \textbf{Model~A'} where, instead of keeping the $N$
best individuals at each generation, we keep $N$ individuals randomly
chosen among the $(3/2)N$ best ones. This allows us to check that our
results remain unchanged under a less stringent selection.

In \textbf{Model B}, each individual~$i$ has infinitely many
offsprings, with positions distributed according to a Poisson point
process of density~$\psi(x-x_i)$ (\textit{i.e.}, with probability
$\psi(x-x_i)\,dx$, there is an offspring of individual~$i$ at
position~$x$). As in Model~A, we only keep the $N$ best offsprings. Here,
$\psi(\epsilon)$ is a positive function such that
$\int\psi(\epsilon)d\epsilon=\infty$ (for the population not to
disappear) and which decays fast enough as $\epsilon\to\infty$
to ensure that these best offsprings have finite positions. Having infinitely many
offsprings at the first step is of course unrealistic, but after selection,
each individual has only a finite number of offsprings. The main two
reasons for considering Model~B are to check the robustness of our
results and to exhibit an exactly soluble case for one particular 
$\psi(\epsilon)$.

\medskip

The genealogical tree of an evolving population can be characterized in
many ways \cite{Schweinsberg.00,Pitman.99}. Here we measure average
coalescence times $\langle T_p\rangle$ defined as follows: $T_p$ is the
age of the most recent common ancestor of $p$ individuals chosen at
random at generation $g$, and $\langle T_p \rangle $ is the average of
$T_p$ over all choices of these $p$ individuals and over all generations
$g$.

In absence of selection (for example when each individual has $k$
offsprings as in model~A, but with $N$ survivors chosen uniformly among 
these $kN$
offsprings), these $\langle T_p \rangle$ grow
linearly with $N$ and their ratios take, for large $N$, the
simple values (independent of $k$) of
the Kingman coalescent \cite{Kingman.82,TavareBGD.97} :
\begin{equation}
\langle T_p \rangle \sim N ,
\ \ 
{\langle T_3 \rangle \over \langle T_2 \rangle} \to {4 \over 3}
, \ \ 
{\langle T_4 \rangle \over \langle T_2 \rangle} \to {3 \over 2}
, \ \ 
{\langle T_p \rangle \over \langle T_2 \rangle} \to 2-{2 \over p}
\label{neutral}
\end{equation}

One goal of the present work is to show that the effect of
selection changes completely Eq.~(\ref{neutral}): the time scale of these
coalescence times $\langle T_p \rangle$ becomes
\begin{equation}
\langle T_p \rangle \sim \big[ \ln N \big] ^\alpha 
\label{alpha}
\end{equation}
and the ratios are compatible with the values characterizing
the Bolthausen-Sznitman
coalescent \cite{Pitman.99,BolthausenSznitman.98,Ruelle.87}:
\begin{equation}
{\langle T_3 \rangle \over \langle T_2 \rangle} \simeq {5 \over 4}
\ \ \ ; \ \ \
{\langle T_4 \rangle \over \langle T_2 \rangle} \simeq {25 \over 18}
\label{selection}
\end{equation}

In Figs.~\ref{fig:T2} and~\ref{fig:ratio}, we show the results of
simulations for different cases :
\begin{figure}[t]
\includegraphics[width=\columnwidth]{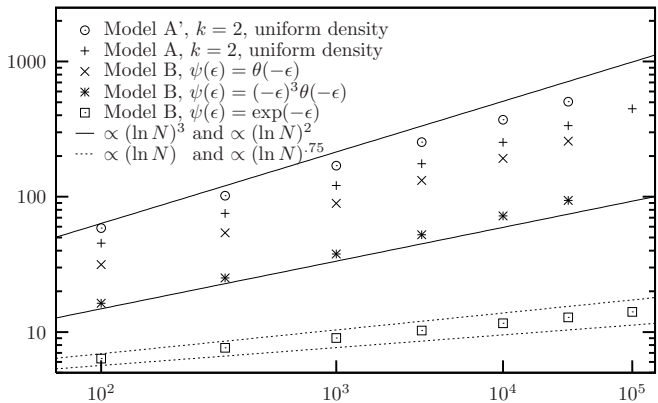}
\caption{Times $\langle T_2\rangle$ versus $N$, for 
different models. The scale of the $N$ axis is $\ln\ln N$. The data
points are compared to several power laws of $\ln N$ shown by the
straight lines.}
\label{fig:T2}
\end{figure}
models A and A' with $k=2$ and a uniform density $\rho(\epsilon)=1$ for
$0<\epsilon<1$, and model~B for three different choices of the density
$\psi(\epsilon)$:
$\psi_1(\epsilon) = \theta(-\epsilon)$,
$\psi_2(\epsilon) = (-\epsilon)^3 \theta(-\epsilon)$ and
$\psi_3(\epsilon) = e^{-\epsilon}$.

Typically we simulated populations of sizes ranging from $N=10^2$ to $10^5$ over
$10^7$ generations.
We measured the times $\langle T_p\rangle$ by recording at each generation $g$
the age $T_2(i,j)$ of the most recent common ancestor of
individuals $i$ and $j$. One then gets $\langle T_2 \rangle$ by averaging
$T_2(i,j)$ over $i$, $j$ and $g$. As the matrix $T_2(i,j)$ is ultrametric, no
additional
information is needed to compute the $\langle T_p\rangle$: for instance,
$T_3(i,j,k)=\max[T_2(i,j), T_2(i,k)]$. For large sizes~$N$, we actually
took advantage of ultrametricity by representing the matrix $T_2(i,j)$ as a tree:
at each step, we only kept track of the
current $N$ individuals and of all the most recent common ancestors of
any pair of them. There are at most $N-1$ such ancestors, so both memory
and execution time grow linearly with $N$, instead of $N^2$ if we
were manipulating the full matrix.

\begin{figure}[t]
\includegraphics[width=\columnwidth]{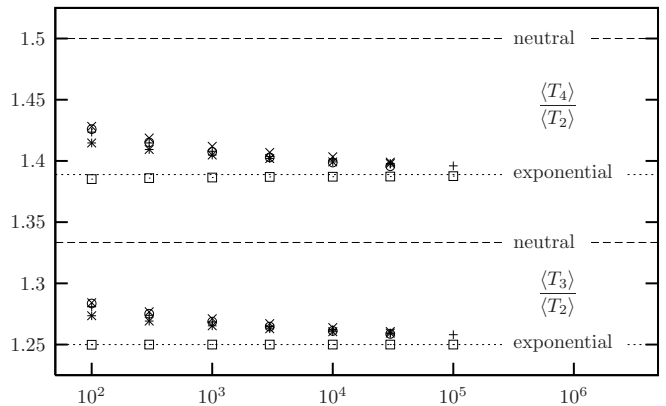}
\caption{Ratios $\langle T_3\rangle/\langle T_2\rangle$ and $\langle
T_4\rangle/\langle T_2\rangle$ versus $N$, for
the same models (same symbols) as in Fig.~\ref{fig:T2}.
The dashed lines represent the the neutral case Eq.~(\ref{neutral}), and
the dotted lines correspond to Eq.~(\ref{selection}), \textit{i.e.}
model B with $\psi(\epsilon)=e^{-\epsilon}$.}
\label{fig:ratio}
\end{figure}

In all cases except for model B with the exponential distribution, which
is special as we shall see below, Fig.~\ref{fig:T2} indicates
that the exponent $\alpha$ defined in Eq.~(\ref{alpha}) is in the range 
\begin{equation}
2. \leq \alpha_{\rm measured} \leq 3.
\label{alpha-measured}
\end{equation} 
For model B with the
exponential distribution, however, our
data suggest a significantly smaller value $.75 \leq \alpha \leq 1 $.

Fig.~\ref{fig:ratio} shows the ratios $\langle T_3 \rangle / \langle T_2 \rangle$ and
$\langle T_4 \rangle / \langle T_2 \rangle$ for the same models as
in Fig.~\ref{fig:T2}.
In all cases, including
model B with $\psi(\epsilon)=e^{-\epsilon}$, these ratios take for large
$N$ 
values similar to Eq.~(\ref{selection}) which differ noticeably from their
values~(\ref{neutral}) in absence of selection. 
At present,
we do not have a general theory to explain this numerical result.

Only for model B with $\psi(\epsilon) = e^{- \epsilon}$ (the special
case), one can calculate these ratios exactly using a method similar to
 \cite{BrunetDerrida.04}. If at generation $g$, the population consists of
$N$ individuals $x_1(g),\dots, x_N(g)$, the probability
of having one of their offsprings in the interval $y, y+dy$ is
$$ \sum_{i=1}^N \psi\big(y-x_i(g)\big) \ dy = 
\sum_{i=1}^N e^{x_i(g) -y} \ dy= 
e^{X_g -y} \ dy$$ where
$$ X_g= \ln \left[ e^{x_1(g)} +e^{x_2(g)}+ \dots + e^{x_N(g)} \right]$$
So, for model B with $\psi(\epsilon)= e^{-\epsilon}$, the offsprings of
the whole population are the same as those of a single effective
individual at position $X_g$.
This means that one can write 
$x_i(g+1)= X_g + y_i$, where 
$y_1,y_2,...,y_N$ are the $N$ largest values of a Poisson point process on
the line with exponential density. The $y_i$'s
are therefore distributed according to
\begin{equation}
{\rm P}(y_N < y_{N-1} < \dots < y_1)= e^{ -(y_1 + y_2 + \dots+
y_N) - e^{-y_N} } 
\label{prob}
\end{equation}
With this simplification, one can see that the differences $\Delta X_g
=X_{g+1} - X_g$ are independent identically distributed random variables.
At generation $g$, the $N$ numbers $x_i(g)$ form a cloud of points
which does not spread in time, very much like a quantum $N$-particle bound
state. This cloud has a well defined velocity $v_N$ and
diffusion constant $D_N$. As the differences $\Delta X_g$ are
independent, $v_N$ and $D_N$ are given by:
$$ v_N = \langle \Delta X_{g} \rangle \ \ \ ; \ \ \
D_N = \langle [\Delta X_g]^2 \rangle - \langle \Delta X_g
\rangle^2 $$
where the expectations are over the distribution (\ref{prob}) of the
$y_i$'s and $$ \Delta X_g =X_{g+1}
- X_g = \ln \left[ e^{y_1}+ ... +e^{y_n}\right]$$
Calculations similar to those of  \cite{BrunetDerrida.04} 
lead for large $N$ to
\begin{equation}
v_N = \ln\ln N + {\ln\ln N + 1 \over \ln N} + \cdots;  \
D_N = {\pi^2 \over 3\ln N}  + \cdots
\label{vNDN}
\end{equation}

We now turn to computing $\langle T_p\rangle$ in the exponential
case. If one chooses randomly $p \geq 2$ individuals 
at generation $g+1$, one can calculate the
probability $q_p$ that they have the same ancestor at generation $g$
\begin{equation}
q_p
=\left\langle {\sum_i e^{p y_i} \over [\sum_i e^{ y_i}]^p} \right\rangle
\simeq {1 \over p-1}{1 \over \ln N} + \cdots
\label{qp}
\end{equation}
One can also show that for large $N$ and fixed $p$, events with
more than one coalescence within the $p$ individuals
between two successive generations have a
probability of order $(\ln N)^{-2}$ at most. Thus, for large $N$,
the genealogical tree of a sample of $p$ individuals consists of single
coalescence events separated by times of order $\ln N$.

From the knowledge of the $q_p$'s, one can obtain for large $N$ the
probability $ r_p(k)$ that $p$ individuals at generation $g+1$ have
exactly $k<p$ ancestors at generation $g$:
\begin{equation}
\begin{aligned}
r_p(k)&= \sum_{j=0}^{k-1} {(-)^{j-k+1} p! \over j! (k-1-j)! (p-k+1)!}
q_{p-j}\\
&= {p \over (p-k)(p-k+1)} {1 \over \ln N} + \cdots
\end{aligned}
\label{rpk}
\end{equation}
and one has
\begin{equation}
\langle T_p \rangle =1 + \langle T_p \rangle + \sum_{k<p} r_p(k)
\ [\langle T_k \rangle - \langle T_p \rangle]
\label{Tprec}
\end{equation}
Using the fact that $\langle T_1 \rangle=0$
this immediately gives 
\begin{equation}
\langle T_2 \rangle \simeq \ln N
\label{T2exp}
\end{equation}
(in reasonable agreement with our simulations of Fig.~\ref{fig:T2}) and all the ratios
$\langle T_n \rangle / \langle T_2 \rangle$.
For $n=3$ and $4$, this gives Eq.~(\ref{selection}) which is
therefore asymptotically exact in the exponential case.

We now return to the general case. Our models are branching processes
which
are known to be
related to fronts of the F-KPP type \cite{McKean.75}. Let us now see how one can
associate to model~B a noisy traveling wave equation (a similar
calculation can be done for model~A). At generation $g$ the whole
population can be characterized by a function $h_g(x)$ which counts
the fraction of individuals $i$ such that $x_i(g) > x$. Obviously
$h_g(x)$ has the shape of a front ($h_g(-\infty)=1$ and $h_g(\infty)=0$).
From the definition of model~B, one can show that $h_g(x)$ satisfies
\begin{equation}
h_{g+1}(x)= \min\left[1, \int h_g(x-\epsilon) \psi(\epsilon) d \epsilon
+ { \eta_g(x) \over \sqrt{N}} \right]
\label{dyn}
\end{equation}
where $\eta_g(x)$ is some correlated noise of
zero-mean with a variance
equal to the integral appearing in
Eq.~(\ref{dyn}). One can show as in  \cite{BrunetDerrida.04} that
this noise is Gaussian far from the tip of the front.

In the large $N$ limit, Eq.~(\ref{dyn}) becomes deterministic.
One can look for traveling wave solutions moving at a velocity
$v$. In the region where $h\ll1$,
an exponential shape $h_g(x) \simeq A
\exp[ - \gamma(x-v g) ]$ is a solution if
\begin{equation}
v=v(\gamma)={1 \over \gamma}\ln\left[ \int e^{\gamma \epsilon}
\psi(\epsilon) d \epsilon\right]
\end{equation}
If the front is of the F-KPP type \cite{vanSaarloos.03}, its velocity for
steep enough initial conditions is the smallest one for which $\gamma$ is
real. Furthermore, for finite but large $N$, one
expects \cite{BrunetDerrida.01,Moro2.04,BDMM.06} a correction to this
velocity of order $[\ln N]^{-2}$ and a diffusion constant scaling like
$[\ln N]^{-3}$. We have checked that these predictions are compatible
with our numerical simulations for all the models that we tested when
$v(\gamma)$ has a minimum value (all the models of
Figs.~\ref{fig:T2},~\ref{fig:ratio} except
the exponential case have this property). For other choices such
as $\psi(\epsilon)=\exp(-\epsilon)$, however, $v(\gamma)=\infty$ for all
$\gamma$, and the front is not of the F-KPP type.
It is therefore not surprising that the genealogy of
the exponential case is special.

In all cases, the times $\langle T_2\rangle$
or $\langle T_p\rangle$ give the order of magnitude of the age of the
most recent common ancestor of the whole population and, therefore, the
time scale on which the population looses memory about its genealogy.
Now, the position of this most recent common ancestor has
fluctuations of order $1$, and this contributes a random shift of order 1 to
the displacement of front. Therefore, the fluctuating part of the
position at generation $g$ is the sum of roughly $g/\langle T_2 \rangle$
independent random variables of order $1$. Within this picture,
the diffusion constant but also all the cumulants of the position would
scale like
\begin{equation}
D_N \sim {1/\langle T_2\rangle }\sim{ 1/\langle T_p\rangle } \ .
\label{diffusion}
\end{equation}
That all cumulants have the same large $N$ dependence was indeed one of
the main results of our previous work \cite{BDMM.06}.
Note that
Eq.~(\ref{diffusion}) does hold (Eqs.~(\ref{vNDN})
and~(\ref{T2exp}))
in the exponential case.
Assuming that it remains valid in general, we would then predict
\begin{equation}
\alpha_\text{prediction}=3
\label{prediction}
\end{equation}
since there is now good numerical
evidence \cite{BrunetDerrida.01,Moro2.04} as well as analytic
arguments \cite{BDMM.06} in favor of $D_N \simeq [\ln N ]^{-3}$ in the
generic case, \textit{i.e.} if $v(\gamma)$ has a finite minimum. We
think that this prediction agrees with our measurements
(\ref{alpha-measured}) up to finite size corrections: in fact, for the
diffusion constant itself, it already
turned out \cite{BrunetDerrida.01,BDMM.06,Moro2.04} that the large $N$
asymptotic regime was only observed for much larger systems than the ones
studied here.

Beyond the fact that the time scale is logarithmic in $N$, which is
not surprising for models of evolution in presence of strong
selection \cite{KaplanHudsonLangley.89,DurrettSchweinsberg.04,DurrettSchweinsberg.05}, the times
$\langle T_p\rangle$ are characteristic of the statistical properties of
the genalogical trees of samples of a few individuals. For the
exponential model, these statistics are totally specified by the fact that
there is at most one single coalescence event at each generation with
probability Eq.~(\ref{qp}). Surprisingly, these coalescence probabilities (up
to the factor $\ln N$ which fixes the time-scale) are the same as those
which emerged from the theory of spin-glasses \cite{BolthausenSznitman.98,Ruelle.87,Pitman.99}, so
that trees, in the exponential model~B here, have exactly the same
statistics as the ultrametric trees of Parisi's mean-field theory of
spin-glasses \cite{Parisi.80,MezardPSTV.84}. So far we have not been able to develop a replica
approach for noisy traveling waves to justify this connection. There
is however some hope to do so since noisy traveling waves appear in the
study of directed polymers in a
random medium \cite{BrunetDerrida.04}, a system for which Parisi's theory
is known to be valid at the mean-field level \cite{DerridaSpohn.88}.

The numerical data presented in this paper show that, for the class of
models we considered, selection has drastic effects on the genealogies:
the coalescence times become logarithmic in the population size
(\ref{alpha}) instead of linear and the statistics of the coalesence
times are modified. The accuracy of our simulations is not sufficient to
be sure that the exponent $\alpha$ and the ratios of coalescence
times are universal (for all models for which $v(\gamma)$ has a
finite minimum). We however gave an argument (\ref{diffusion})  which
supports the conjecture (\ref{prediction}). Of course,
developing an
analytical approach susceptible of proving or disproving this
universality is a challenging open problem. Another open issue is whether
thinking in terms of genealogies is limited to the family of selection
models discussed here or could be extended to more general noisy
traveling wave equations.

Lastly, it would be interesting to know what our ratios
(\ref{selection}) would become in other models of
evolution with selection such as  \cite{Tsimring.96,Kessler.97}
and if there is a chance of estimating them from
experimental data on genetic diversity.

\medskip

This work was partially supported by the U.S. Department of Energy.


\begin{thebibliography}{41}
\expandafter\ifx\csname natexlab\endcsname\relax\def\natexlab#1{#1}\fi
\expandafter\ifx\csname bibnamefont\endcsname\relax
  \def\bibnamefont#1{#1}\fi
\expandafter\ifx\csname bibfnamefont\endcsname\relax
  \def\bibfnamefont#1{#1}\fi
\expandafter\ifx\csname citenamefont\endcsname\relax
  \def\citenamefont#1{#1}\fi
\expandafter\ifx\csname url\endcsname\relax
  \def\url#1{\texttt{#1}}\fi
\expandafter\ifx\csname urlprefix\endcsname\relax\def\urlprefix{URL }\fi
\providecommand{\bibinfo}[2]{#2}
\providecommand{\eprint}[2][]{\url{#2}}

\bibitem[{\citenamefont{Fisher}(1937)}]{Fisher.37}
\bibinfo{author}{\bibfnamefont{R.~A.} \bibnamefont{Fisher}},
  \bibinfo{journal}{Annals of Eugenics} \textbf{\bibinfo{volume}{7}},
  \bibinfo{pages}{355} (\bibinfo{year}{1937}).

\bibitem[{\citenamefont{Kolmogorov et~al.}(1937)\citenamefont{Kolmogorov,
  Petrovsky, and Piscounov}}]{KPP.37}
\bibinfo{author}{\bibfnamefont{A.}~\bibnamefont{Kolmogorov}},
  \bibinfo{author}{\bibfnamefont{I.}~\bibnamefont{Petrovsky}},
  \bibnamefont{and}
  \bibinfo{author}{\bibfnamefont{N.}~\bibnamefont{Piscounov}},
  \bibinfo{journal}{Bull. Univ. \'Etat Moscou, A} \textbf{\bibinfo{volume}{1}},
  \bibinfo{pages}{1} (\bibinfo{year}{1937}).

\bibitem[{\citenamefont{van Saarloos}(2003)}]{vanSaarloos.03}
\bibinfo{author}{\bibfnamefont{W.}~\bibnamefont{van Saarloos}},
  \bibinfo{journal}{Phys. Rep.} \textbf{\bibinfo{volume}{386}},
  \bibinfo{pages}{29} (\bibinfo{year}{2003}).

\bibitem[{\citenamefont{Derrida and Spohn}(1988)}]{DerridaSpohn.88}
\bibinfo{author}{\bibfnamefont{B.}~\bibnamefont{Derrida}} \bibnamefont{and}
  \bibinfo{author}{\bibfnamefont{H.}~\bibnamefont{Spohn}}, \bibinfo{journal}{J.
  Stat. Phys.} \textbf{\bibinfo{volume}{51}}, \bibinfo{pages}{817}
  (\bibinfo{year}{1988}).

\bibitem[{\citenamefont{Brunet and Derrida}(2004)}]{BrunetDerrida.04}
\bibinfo{author}{\bibfnamefont{{\'E}.}~\bibnamefont{Brunet}} \bibnamefont{and}
  \bibinfo{author}{\bibfnamefont{B.}~\bibnamefont{Derrida}},
  \bibinfo{journal}{Phys. Rev. E} \textbf{\bibinfo{volume}{70}},
  \bibinfo{pages}{016106} (\bibinfo{year}{2004}).

\bibitem[{\citenamefont{Carpentier and {Le
  Doussal}}(2000)}]{CarpentierLedoussal.00}
\bibinfo{author}{\bibfnamefont{D.}~\bibnamefont{Carpentier}} \bibnamefont{and}
  \bibinfo{author}{\bibfnamefont{P.}~\bibnamefont{{Le Doussal}}},
  \bibinfo{journal}{Nucl. Phys. B} \textbf{\bibinfo{volume}{588}},
  \bibinfo{pages}{531} (\bibinfo{year}{2000}).

\bibitem[{\citenamefont{Iancu et~al.}(2005)\citenamefont{Iancu, Mueller, and
  Munier}}]{IancuMuellerMunier.05}
\bibinfo{author}{\bibfnamefont{E.}~\bibnamefont{Iancu}},
  \bibinfo{author}{\bibfnamefont{A.~H.} \bibnamefont{Mueller}},
  \bibnamefont{and} \bibinfo{author}{\bibfnamefont{S.}~\bibnamefont{Munier}},
  \bibinfo{journal}{Phys. Lett. B} \textbf{\bibinfo{volume}{606}},
  \bibinfo{pages}{342} (\bibinfo{year}{2005}).

\bibitem[{\citenamefont{Breuer et~al.}(1995)\citenamefont{Breuer, Huber, and
  Petruccione}}]{Breuer.95}
\bibinfo{author}{\bibfnamefont{H.-P.} \bibnamefont{Breuer}},
  \bibinfo{author}{\bibfnamefont{W.}~\bibnamefont{Huber}}, \bibnamefont{and}
  \bibinfo{author}{\bibfnamefont{F.}~\bibnamefont{Petruccione}},
  \bibinfo{journal}{Europhysics Letters} \textbf{\bibinfo{volume}{30}},
  \bibinfo{pages}{69} (\bibinfo{year}{1995}).

\bibitem[{\citenamefont{Doering et~al.}(2003)\citenamefont{Doering, Mueller,
  and Smereka}}]{DoeringMuellerSmereka.03}
\bibinfo{author}{\bibfnamefont{C.~R.} \bibnamefont{Doering}},
  \bibinfo{author}{\bibfnamefont{C.}~\bibnamefont{Mueller}}, \bibnamefont{and}
  \bibinfo{author}{\bibfnamefont{P.}~\bibnamefont{Smereka}},
  \bibinfo{journal}{Physica A} \textbf{\bibinfo{volume}{325}},
  \bibinfo{pages}{243} (\bibinfo{year}{2003}).

\bibitem[{\citenamefont{Krapivsky and Majumdar}(2000)}]{KrapivskyMajumdar.00}
\bibinfo{author}{\bibfnamefont{P.~L.} \bibnamefont{Krapivsky}}
  \bibnamefont{and} \bibinfo{author}{\bibfnamefont{S.~N.}
  \bibnamefont{Majumdar}}, \bibinfo{journal}{Phys. Rev. Lett.}
  \textbf{\bibinfo{volume}{85}}, \bibinfo{pages}{5492} (\bibinfo{year}{2000}).

\bibitem[{\citenamefont{Lemarchand et~al.}(1995)\citenamefont{Lemarchand,
  Lesne, and Mareschal}}]{Lemarchand.95}
\bibinfo{author}{\bibfnamefont{A.}~\bibnamefont{Lemarchand}},
  \bibinfo{author}{\bibfnamefont{A.}~\bibnamefont{Lesne}}, \bibnamefont{and}
  \bibinfo{author}{\bibfnamefont{M.}~\bibnamefont{Mareschal}},
  \bibinfo{journal}{Phys. Rev. E} \textbf{\bibinfo{volume}{51}},
  \bibinfo{pages}{4457} (\bibinfo{year}{1995}).

\bibitem[{\citenamefont{Mueller and Sowers}(1995)}]{MuellerSowers.95}
\bibinfo{author}{\bibfnamefont{C.}~\bibnamefont{Mueller}} \bibnamefont{and}
  \bibinfo{author}{\bibfnamefont{R.~B.} \bibnamefont{Sowers}},
  \bibinfo{journal}{J. Funct. Anal.} \textbf{\bibinfo{volume}{128}},
  \bibinfo{pages}{439} (\bibinfo{year}{1995}).

\bibitem[{\citenamefont{Mai et~al.}(1996)\citenamefont{Mai, Sokolov, and
  Blumen}}]{Mai.96}
\bibinfo{author}{\bibfnamefont{J.}~\bibnamefont{Mai}},
  \bibinfo{author}{\bibfnamefont{I.~M.} \bibnamefont{Sokolov}},
  \bibnamefont{and} \bibinfo{author}{\bibfnamefont{A.}~\bibnamefont{Blumen}},
  \bibinfo{journal}{Phys. Rev. Lett.} \textbf{\bibinfo{volume}{77}},
  \bibinfo{pages}{4462} (\bibinfo{year}{1996}).

\bibitem[{\citenamefont{Panja}(2004)}]{Panja.03}
\bibinfo{author}{\bibfnamefont{D.}~\bibnamefont{Panja}},
  \bibinfo{journal}{Phys. Rep.} \textbf{\bibinfo{volume}{393}},
  \bibinfo{pages}{87} (\bibinfo{year}{2004}).

\bibitem[{\citenamefont{Brunet and Derrida}(2001)}]{BrunetDerrida.01}
\bibinfo{author}{\bibfnamefont{{\'E}.}~\bibnamefont{Brunet}} \bibnamefont{and}
  \bibinfo{author}{\bibfnamefont{B.}~\bibnamefont{Derrida}},
  \bibinfo{journal}{J. Stat. Phys.} \textbf{\bibinfo{volume}{103}},
  \bibinfo{pages}{269} (\bibinfo{year}{2001}).

\bibitem[{\citenamefont{Kloster}(2005)}]{Kloster.05}
\bibinfo{author}{\bibfnamefont{M.}~\bibnamefont{Kloster}},
  \bibinfo{journal}{Phys. Rev. Lett.} \textbf{\bibinfo{volume}{95}},
  \bibinfo{pages}{168701} (\bibinfo{year}{2005}).

\bibitem[{\citenamefont{Brunet et~al.}(2006)\citenamefont{Brunet, Derrida,
  Mueller, and Munier}}]{BDMM.06}
\bibinfo{author}{\bibfnamefont{E.}~\bibnamefont{Brunet}},
  \bibinfo{author}{\bibfnamefont{B.}~\bibnamefont{Derrida}},
  \bibinfo{author}{\bibfnamefont{A.~H.} \bibnamefont{Mueller}},
  \bibnamefont{and} \bibinfo{author}{\bibfnamefont{S.}~\bibnamefont{Munier}}
  (\bibinfo{year}{2006}), \bibinfo{note}{{Phys. Rev. E}, in press}.

\bibitem[{\citenamefont{Moro}(2004{\natexlab{a}})}]{Moro.04}
\bibinfo{author}{\bibfnamefont{E.}~\bibnamefont{Moro}}, \bibinfo{journal}{Phys.
  Rev. E} \textbf{\bibinfo{volume}{69}}, \bibinfo{pages}{060101(R)}
  (\bibinfo{year}{2004}{\natexlab{a}}).

\bibitem[{\citenamefont{Kessler et~al.}(1998)\citenamefont{Kessler, Ner, and
  Sander}}]{Kessler.98}
\bibinfo{author}{\bibfnamefont{D.~A.} \bibnamefont{Kessler}},
  \bibinfo{author}{\bibfnamefont{Z.}~\bibnamefont{Ner}}, \bibnamefont{and}
  \bibinfo{author}{\bibfnamefont{L.~M.} \bibnamefont{Sander}},
  \bibinfo{journal}{Phys. Rev. E} \textbf{\bibinfo{volume}{58}},
  \bibinfo{pages}{107} (\bibinfo{year}{1998}).

\bibitem[{\citenamefont{Colon and Doering}(2005)}]{ColonDoering.05}
\bibinfo{author}{\bibfnamefont{J.~G.} \bibnamefont{Colon}} \bibnamefont{and}
  \bibinfo{author}{\bibfnamefont{C.~R.} \bibnamefont{Doering}},
  \bibinfo{journal}{J. Stat. Phys.} \textbf{\bibinfo{volume}{120}},
  \bibinfo{pages}{421} (\bibinfo{year}{2005}).

\bibitem[{\citenamefont{Pechenik and Levine}(1999)}]{PechenikLevine.99}
\bibinfo{author}{\bibfnamefont{L.}~\bibnamefont{Pechenik}} \bibnamefont{and}
  \bibinfo{author}{\bibfnamefont{H.}~\bibnamefont{Levine}},
  \bibinfo{journal}{Phys. Rev. E} \textbf{\bibinfo{volume}{59}},
  \bibinfo{pages}{3893} (\bibinfo{year}{1999}).

\bibitem[{\citenamefont{Escudero}(2004)}]{Escudero.04}
\bibinfo{author}{\bibfnamefont{C.}~\bibnamefont{Escudero}},
  \bibinfo{journal}{Phys. Rev. E} \textbf{\bibinfo{volume}{70}},
  \bibinfo{pages}{041102} (\bibinfo{year}{2004}).

\bibitem[{\citenamefont{Brunet and Derrida}(1997)}]{BrunetDerrida.97}
\bibinfo{author}{\bibfnamefont{{\'E}.}~\bibnamefont{Brunet}} \bibnamefont{and}
  \bibinfo{author}{\bibfnamefont{B.}~\bibnamefont{Derrida}},
  \bibinfo{journal}{Phys. Rev. E} \textbf{\bibinfo{volume}{56}},
  \bibinfo{pages}{2597} (\bibinfo{year}{1997}).

\bibitem[{\citenamefont{Peng et~al.}(2003)\citenamefont{Peng, Gerland, Hwa, and
  Levine}}]{PengGerlandHwaLevine.03}
\bibinfo{author}{\bibfnamefont{W.}~\bibnamefont{Peng}},
  \bibinfo{author}{\bibfnamefont{U.}~\bibnamefont{Gerland}},
  \bibinfo{author}{\bibfnamefont{T.}~\bibnamefont{Hwa}}, \bibnamefont{and}
  \bibinfo{author}{\bibfnamefont{H.}~\bibnamefont{Levine}},
  \bibinfo{journal}{Phys. Rev. Lett.} \textbf{\bibinfo{volume}{90}},
  \bibinfo{eid}{088103} (\bibinfo{year}{2003}).

\bibitem[{\citenamefont{Snyder}(2003)}]{Snyder.03}
\bibinfo{author}{\bibfnamefont{R.~E.} \bibnamefont{Snyder}},
  \bibinfo{journal}{Ecol.} \textbf{\bibinfo{volume}{84}}, \bibinfo{pages}{1333}
  (\bibinfo{year}{2003}).

\bibitem[{\citenamefont{Kloster and Tang}(2004)}]{KlosterTang.04}
\bibinfo{author}{\bibfnamefont{M.}~\bibnamefont{Kloster}} \bibnamefont{and}
  \bibinfo{author}{\bibfnamefont{C.}~\bibnamefont{Tang}},
  \bibinfo{journal}{Phys. Rev. Lett.} \textbf{\bibinfo{volume}{92}},
  \bibinfo{pages}{038101} (\bibinfo{year}{2004}).

\bibitem[{\citenamefont{Schweinsberg}(2000)}]{Schweinsberg.00}
\bibinfo{author}{\bibfnamefont{J.}~\bibnamefont{Schweinsberg}},
  \bibinfo{journal}{Elect. Journ. Prob.} \textbf{\bibinfo{volume}{5}},
  \bibinfo{pages}{1} (\bibinfo{year}{2000}).

\bibitem[{\citenamefont{Pitman}(1999)}]{Pitman.99}
\bibinfo{author}{\bibfnamefont{J.}~\bibnamefont{Pitman}},
  \bibinfo{journal}{Ann. Probab.} \textbf{\bibinfo{volume}{27}},
  \bibinfo{pages}{1870} (\bibinfo{year}{1999}).

\bibitem[{\citenamefont{Kingman}(1982)}]{Kingman.82}
\bibinfo{author}{\bibfnamefont{J.~F.~C.} \bibnamefont{Kingman}},
  \bibinfo{journal}{J. Appl. Probab.} \textbf{\bibinfo{volume}{19A}},
  \bibinfo{pages}{27} (\bibinfo{year}{1982}).

\bibitem[{\citenamefont{Tavar\'e et~al.}(1997)\citenamefont{Tavar\'e, Balding,
  Griffiths, and Donelly}}]{TavareBGD.97}
\bibinfo{author}{\bibfnamefont{S.}~\bibnamefont{Tavar\'e}},
  \bibinfo{author}{\bibfnamefont{D.~J.} \bibnamefont{Balding}},
  \bibinfo{author}{\bibfnamefont{R.~C.} \bibnamefont{Griffiths}},
  \bibnamefont{and} \bibinfo{author}{\bibfnamefont{P.}~\bibnamefont{Donelly}},
  \bibinfo{journal}{Genetics} \textbf{\bibinfo{volume}{145}},
  \bibinfo{pages}{505} (\bibinfo{year}{1997}).

\bibitem[{\citenamefont{Bolthausen and Sznitman}(1998)}]{BolthausenSznitman.98}
\bibinfo{author}{\bibfnamefont{E.}~\bibnamefont{Bolthausen}} \bibnamefont{and}
  \bibinfo{author}{\bibfnamefont{A.-S.} \bibnamefont{Sznitman}},
  \bibinfo{journal}{Com. Math. Phys.} \textbf{\bibinfo{volume}{197}},
  \bibinfo{pages}{247} (\bibinfo{year}{1998}).

\bibitem[{\citenamefont{Ruelle}(1987)}]{Ruelle.87}
\bibinfo{author}{\bibfnamefont{D.}~\bibnamefont{Ruelle}},
  \bibinfo{journal}{Com. Math. Phys} \textbf{\bibinfo{volume}{108}},
  \bibinfo{pages}{225} (\bibinfo{year}{1987}).

\bibitem[{\citenamefont{McKean}(1975)}]{McKean.75}
\bibinfo{author}{\bibfnamefont{H.~P.} \bibnamefont{McKean}},
  \bibinfo{journal}{Comm. Pure Appl. Math.} \textbf{\bibinfo{volume}{28}},
  \bibinfo{pages}{323} (\bibinfo{year}{1975}).

\bibitem[{\citenamefont{Moro}(2004{\natexlab{b}})}]{Moro2.04}
\bibinfo{author}{\bibfnamefont{E.}~\bibnamefont{Moro}}, \bibinfo{journal}{Phys.
  Rev. E} \textbf{\bibinfo{volume}{70}}, \bibinfo{pages}{045102(R)}
  (\bibinfo{year}{2004}{\natexlab{b}}).

\bibitem[{\citenamefont{Kaplan et~al.}(1989)\citenamefont{Kaplan, Hudson, and
  Langley}}]{KaplanHudsonLangley.89}
\bibinfo{author}{\bibfnamefont{N.~L.} \bibnamefont{Kaplan}},
  \bibinfo{author}{\bibfnamefont{R.~R.} \bibnamefont{Hudson}},
  \bibnamefont{and} \bibinfo{author}{\bibfnamefont{C.~H.}
  \bibnamefont{Langley}}, \bibinfo{journal}{Genetics}
  \textbf{\bibinfo{volume}{123}}, \bibinfo{pages}{887} (\bibinfo{year}{1989}).

\bibitem[{\citenamefont{Durrett and
  Schweinsberg}(2004)}]{DurrettSchweinsberg.04}
\bibinfo{author}{\bibfnamefont{R.}~\bibnamefont{Durrett}} \bibnamefont{and}
  \bibinfo{author}{\bibfnamefont{J.}~\bibnamefont{Schweinsberg}},
  \bibinfo{journal}{Theor. Population Biol.} \textbf{\bibinfo{volume}{66}},
  \bibinfo{pages}{129} (\bibinfo{year}{2004}).

\bibitem[{\citenamefont{Durrett and
  Schweinsberg}(2005)}]{DurrettSchweinsberg.05}
\bibinfo{author}{\bibfnamefont{R.}~\bibnamefont{Durrett}} \bibnamefont{and}
  \bibinfo{author}{\bibfnamefont{J.}~\bibnamefont{Schweinsberg}},
  \bibinfo{journal}{Stoch. Proc. Appl.} \textbf{\bibinfo{volume}{115}},
  \bibinfo{pages}{1628} (\bibinfo{year}{2005}).

\bibitem[{\citenamefont{Parisi}(1980)}]{Parisi.80}
\bibinfo{author}{\bibfnamefont{G.}~\bibnamefont{Parisi}}, \bibinfo{journal}{J.
  Phys. A} \textbf{\bibinfo{volume}{13}}, \bibinfo{pages}{1101}
  (\bibinfo{year}{1980}).

\bibitem[{\citenamefont{M\'ezard et~al.}(1984)\citenamefont{M\'ezard, Parisi,
  Sourlas, Toulouse, and Virasoro}}]{MezardPSTV.84}
\bibinfo{author}{\bibfnamefont{M.}~\bibnamefont{M\'ezard}},
  \bibinfo{author}{\bibfnamefont{G.}~\bibnamefont{Parisi}},
  \bibinfo{author}{\bibfnamefont{N.}~\bibnamefont{Sourlas}},
  \bibinfo{author}{\bibfnamefont{G.}~\bibnamefont{Toulouse}}, \bibnamefont{and}
  \bibinfo{author}{\bibfnamefont{M.~A.} \bibnamefont{Virasoro}},
  \bibinfo{journal}{Journal de Physique} \textbf{\bibinfo{volume}{45}},
  \bibinfo{pages}{843} (\bibinfo{year}{1984}).

\bibitem[{\citenamefont{Tsimring et~al.}(1996)\citenamefont{Tsimring, Levine,
  and Kessler}}]{Tsimring.96}
\bibinfo{author}{\bibfnamefont{L.}~\bibnamefont{Tsimring}},
  \bibinfo{author}{\bibfnamefont{H.}~\bibnamefont{Levine}}, \bibnamefont{and}
  \bibinfo{author}{\bibfnamefont{D.~A.} \bibnamefont{Kessler}},
  \bibinfo{journal}{Phys. Rev. Lett.} \textbf{\bibinfo{volume}{76}},
  \bibinfo{pages}{4440} (\bibinfo{year}{1996}).

\bibitem[{\citenamefont{Kessler et~al.}(1997)\citenamefont{Kessler, Levine,
  Ridgway, and Tsimring}}]{Kessler.97}
\bibinfo{author}{\bibfnamefont{D.~A.} \bibnamefont{Kessler}},
  \bibinfo{author}{\bibfnamefont{H.}~\bibnamefont{Levine}},
  \bibinfo{author}{\bibfnamefont{D.}~\bibnamefont{Ridgway}}, \bibnamefont{and}
  \bibinfo{author}{\bibfnamefont{L.}~\bibnamefont{Tsimring}},
  \bibinfo{journal}{J. Stat. Phys.} \textbf{\bibinfo{volume}{87}},
  \bibinfo{pages}{519} (\bibinfo{year}{1997}).

\end{thebibliography}
\end{document}